\documentclass[aps,floats,draft,showpacs]{revtex4}
\usepackage{psfig}
\usepackage{amsmath}

\begin{document}
\def\cs{c_{\rm s}}
\def\rhos{\rho_{\rm s}}
\def\rc{r_{\rm c}}
\def\sc{s_{\rm c}}
\def\l{\lambda}

\newcommand{\Gn}
    {\setlength{\fboxsep}{-.4pt}\fbox{\rule{3.0em}{0mm}\rule{0mm}{1ex}}}
\newcommand{\Fn}{\rule{3.0em}{1ex}}

\voffset=2cm

\title{Mean-field solution of the parity-conserving kinetic phase transition
in one dimension}
\author{Dexin Zhong$^1$, Daniel ben-Avraham$^1$, and Miguel A. Mu{\~n}oz$^2$} 
\affiliation{\mbox{Department of Physics, Clarkson University, Postdam, New York
13699-5820 USA 
}}
\affiliation{\mbox{Instituto de F{\'\i}sica Te{\'o}rica 
y Computacional Carlos I. Universidad de Granada, 18071 Granada, Spain}}
\date{\today} 

\begin{abstract}
A two-offspring branching annihilating random walk model, with finite reaction rates, is 
studied in one-dimension. The model exhibits a transition from an active to an
absorbing  phase, expected to belong to the
$DP2$ universality class embracing systems that possess two symmetric absorbing states,
which in one-dimensional systems, is in many cases equivalent to parity conservation.
The phase transition is studied analytically through a mean-field like 
modification of the so-called {\it parity interval method}.  
The original method of parity intervals allows for an exact analysis of
 the diffusion-controlled limit of infinite reaction rate, 
where there is no active phase and hence no phase transition.  
For finite rates, we obtain a surprisingly good
description of the transition which compares favorably with the outcome of Monte
Carlo simulations.
This provides one of the first analytical attempts to deal with the broadly studied DP2
universality class.
\end{abstract} 

\maketitle

  \section{Introduction}

 Phase transitions occurring away from thermodynamical equilibrium constitute 
one of the most challenging topics in statistical physics. 
They appear in a host of physical systems as well as in many
models in biology, chemistry, sociology, etc. 
Given the lack of a general theory of non-equilibrium systems, 
we are still living a {\it taxonomic era} in this field:
it would be highly desirable to reach a complete classification of the 
known phase transitions into universality classes, as a preliminary step to their full
categorization, and identification of relevant features.

After more than twenty years of study, it has become clear 
that the degree of universality is much narrower away from equilibrium than 
it is in equilibrium transitions. 
Certainly, general properties such as symmetries, conservation
laws, dimensionalities, etc., play a key role, as they do in equilibrium. 
But some other ingredients, such as microscopic dynamical details, 
hard core interactions, and the type of updating may, in some cases,
influence the emerging critical behavior of
non-equilibrium systems, making the task of theoreticians
both stimulating and difficult.

   One of the most robust  and best studied non-equilibrium classes of transitions   
is {\it directed percolation} (DP).  It includes an amazing variety of models and systems
exhibiting a transition from an active to an absorbing phase 
\cite{Kin,DK,CP,Reviews,Marro,IAS},
and is well characterized at a field-theoretical level by Reggeon
field theory (RFT) \cite{conjecture,RFT}.

DP is so robust, that identifying and classifying the nature of 
perturbations able to drive absorbing phase transitions away from this class 
has become a challenging task.
Probably the best known instance of this is the $DP2$ class,
where the presence of two perfectly
symmetric absorbing states ($Z_2$ symmetry) is the main 
feature responsible for non-DP scaling \cite{equiv}.
This class includes a cellular 
automaton introduced by Grassberger et al.,~\cite{peter},
interacting monomer-dimer models \cite{Park}, non-equilibrium
Ising models \cite{Menyhard,Odor}, monomer-monomer surface reaction models \cite{Brown},
$Z_2$-symmetric generalizations of DP \cite{Hinrichsen} 
and of the contact process \cite{Inui},  branching annihilating random walks 
with conserved parity  \cite{TT,Santos,Bramson},  and it has also been related to
generalized versions of the Voter model \cite{Voter1,Voter2}. 
Extensive numerical simulations led
to conjecture rational exponent values for the DP2 class \cite{Jensen}, though this
was later disproved by more exhaustive simulations \cite{Reviews}.

   Although the existence and robustness of the DP2 class are 
well established from a numerical view-point, a solid theoretical understanding 
is still missing. Bold attempts to write down and renormalize a field theory suitable
for the DP2 class have been performed, but the results are not as satisfactory as they
are for RFT \cite{tauber,Korea}. 
In particular, the renormalization is based on a clever, but somehow 
uncontrolled expansion around two-different critical dimensions. 
  As we shall illustrate below,
straightforward mean-field approaches (cluster approximations) fail to 
reproduce a phase transition if too small clusters are considered, and one
has to resort to large-clusters, which make the calculation complicated and
not very accurate, i.e. the convergence of the series is very slow 
(although results can be improved if combined with 
{\it coherent-anomaly methods} \cite{Odor}).

  It is the purpose of this paper to shed some light on these issues, by
examining  models  of Branching Annihilating Walks  with two offspring (2-BAW)
 in one dimension:
the 2-BAW with {\it finite} reaction rate exhibits a phase transition in the DP2
 class \cite{2-BAW}.
Our approach is based on a mean-field modification of the {\it method of parity intervals}, 
originally introduced for the study of annihilation reactions, $A+A\to0$ 
\cite{book,parity_ints,masser,lind}.
A similar approach has been previously employed in conjunction with the
{\it method of empty intervals} \cite{book,ddb,Henkel} for the analysis of other
intractable models that do not conserve parity \cite{approx}.
See \cite{parity_ints,masser} for a more detailed introduction to this method and its 
applications to different models.

The paper is organized as follows. The 2-BAW model is described
 in Section~\ref{model}, where we also perform simple (up to two sites) 
cluster approximations (which fail to capture the transition). 
In Section~\ref{exact} we present the exact solution
of the 2-BAW model in the limit of {\it infinite} reaction rate, using the method
of parity-intervals.  Although there is no phase transition in this limit, the
analysis serves as a basis for the mean-field like approximation presented in
Section~\ref{approxi}, for the relevant case of finite reaction rates.  There we derive
the approximate steady-state solution for generic parameter values, and compare it
with the outcome of Monte Carlo computer simulations.  We conclude with a critical
discussion of our results and further developments, in Section~\ref{disc}.

\section{The model and cluster approximations
\label{model}}
 
 The model is defined as follows. Each site of a one-dimensional lattice
is either empty or singly occupied. The lattice is updated asynchronously:
an occupied site is randomly chosen, and it is tried for diffusion, 
at rate $\Gamma$ (probability $\Gamma/(\Gamma+\Omega)$), 
or branching, at rate $\Omega$ (probability  $\Omega/(\Gamma+\Omega)$); 
time is increased by $1/N$, where $N$ is the number of occupied (active) sites. 
In a diffusion attempt the particle is moved to one of its two nearest neighbors
 (target sites), with equal probabilities.  If the target site is occupied,
 annihilation results with probability $r$: the move is effected and both
 particles, the diffuser and the target, are removed from the lattice. 
 The move is rejected with probability $1-r$, and the lattice state remains unchanged.  
In a branching attempt the particle gives birth to two new particles onto
 the nearest neighbor sites (target sites).  If either, or both of the
 target sites is occupied, annihilation takes place with probability $r$:
 branching is effected, and the occupied target site(s) become empty.
  Once again, the move is rejected with probability $1-r$, and the
 lattice state remains unchanged.

In the diffusion-controlled limit of infinite reaction rate ($r=1$),
 the 2-BAW model is known to possess only a steady absorbing phase \cite{Sudbury}. 
 However, for finite rates ($r<1$), a transition belonging to the DP2 class,
 from an absorbing state, at $r<\rc$, to an active phase, at $r>\rc$, is
 found in numerical simulations~\cite{2-BAW}.  We now demonstrate that
 simple mean-field approximations fail to capture this transition.

The simplest conceivable mean-field theory is that of the one-site
approximation, obtained by neglecting all correlations
between the states of different sites. For example, if the
probability of one site being occupied is equal to the
concentration $c$, then the probability of finding one occupied
site followed immediately by an empty site is $c(1-c)$. There are only
three events that lead to a change in particle concentration:
({\bf i})~annihilation of two particles by diffusion,
$\bullet\bullet\to\circ\circ$;  ({\bf ii})~creation of two particles via birth,
$\circ\bullet\circ\to\bullet\bullet\bullet$; and ({\bf iii})~annihilation of two
particles due to birth onto previously occupied sites,
$\bullet\bullet\bullet\to\circ\bullet\circ$.  Note that the birth process
$\circ\bullet\bullet\to\bullet\bullet\circ$ (and its mirror-symmetric
image) does not alter the concentration.  However, processes (i) and (iii) occur with
restricted probability $r$.  Thus
\begin{equation}
\label{cdot1site}
\frac{d}{dt}c=-2rc^2+2c(1-c)^2-2rc^3\;,
\end{equation}
where we took $\Gamma=\Omega=1$.  The first, second and third terms
 on the r.h.s.\ correspond to ({\bf i}), ({\bf ii}) and ({\bf iii}), respectively.

\begin{figure}
\centerline{\psfig{file=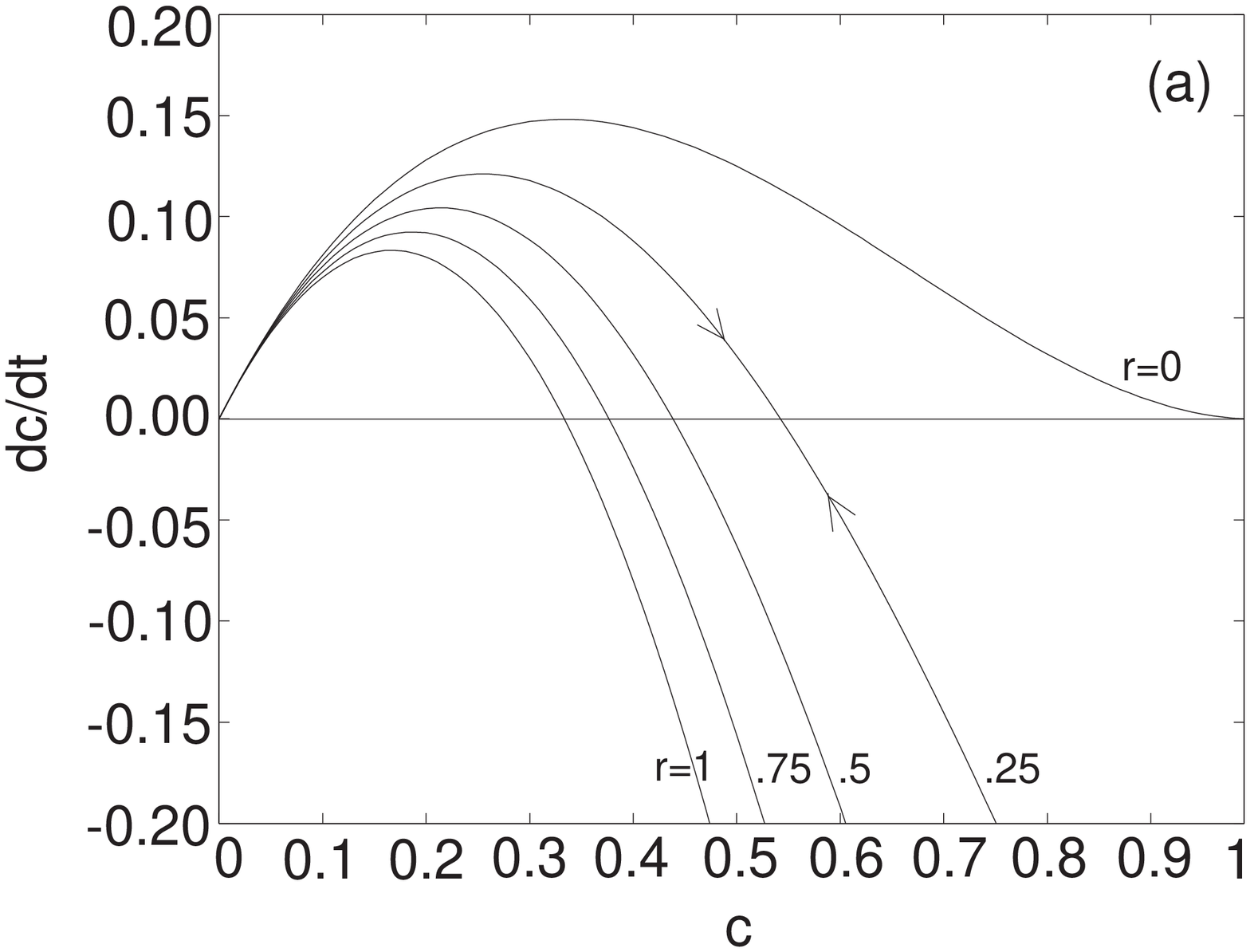,width=7cm,angle=0}} 
\vspace{0.2cm}
\caption{One-site cluster approximation. Flow diagram, according to
Eq.~(\ref{cdot1site}), with $dc/dt=0$.  This shows that the root $\cs=0$
is unstable, while the root $\cs>0$ is stable. Results from the two-site approximation
are very similar. The stationary solution as a function of $r$ is plotted in
Fig.2} 
\label{cs.fig}
\end{figure}

For the steady state, we set $dc/dt=0$
and choose the stable root (Fig.~\ref{cs.fig}), $\cs= (2 + r - \sqrt{8r +
r^2})/{(2-2r)}$.  We conclude that according to the one-site approximation
the system always evolves to an active phase, with $1/3<\cs<1$ (Fig.2).  

The two-site approximation provides the next level of complexity.  Let $p$
be the conditional probability for a site to be occupied, given that the
adjacent site (to its left) is occupied.  Let $q$ be the conditional
probability for a site to be occupied, given that the site to its left is
empty.  Thus, the probability of the events $\bullet\bullet$,
$\bullet\circ$, $\circ\bullet$, $\circ\circ$ is $cp$, $c(1-p)$, $(1-c)q$,
$(1-c)(1-q)$, respectively.  Note that since
$\bullet\circ$ and $\circ\bullet$ are equally likely,
we have,
\begin{equation}
\label{p-q}
c(1-p)=(1-c)q\;.
\end{equation}
(The same relation may be derived from the fact that
$\Pr(\bullet)=\Pr(\bullet\bullet)+\Pr(\circ\bullet)$.)
Eq.~(\ref{cdot1site}) is now rewritten as
\begin{equation}
\label{cdot2site}
\frac{d}{dt}c=-2rcp+2(1-c)q(1-p)-2rcp^2\;,
\end{equation}
where the terms on the r.h.s.\ correspond to the same events as before. 
Setting $dc/dt=0$ and using the relation~(\ref{p-q}), we obtain, for the steady
state,
\[
0=-2rcp+2c(1-p)^2-2rcp^2\;.
\]
We conclude that either $\cs=0$ or $p_{\rm s}=(2 + r - \sqrt{8r +
r^2})/{(2-2r)}$.  In the latter case, an additional evolution equation, for
the event
$\bullet\bullet$, provides the missing value of $\cs$: $\cs>0$ for all
$r$.  The result agrees closely with that of the one-site approximation,
 with deviations not larger than 20$\,$\%.  Thus, also the two-site 
approximation fails to predict the
transition. 
Considering larger clusters, a phase transition can be generated 
\cite{Odor,Attila}, but a large set of coupled equations (which enlarges
with cluster-size) has to be solved, the accuracy is not good, 
and the results are expected to be valid only up to the cluster-size. 
In the next section we present an alternative method intended to overcome 
these difficulties.

\section{Parity intervals and exact solution for $r=1$ \label{exact}}

We now turn to a different approach, that of the method of parity
intervals~\cite{book,parity_ints,masser,lind}.  We first present the
 exactly soluble case of $r=1$, for which there is no transition.
The exact approach followed here serves as a basis for a mean-field
 like approximation, for the more interesting case of $r<1$
--- an approximation that does capture the transition and reproduces
kinetic details surprisingly well.

Let $G_n(t)$ be the probability that (in an homogeneous system) an
arbitrary segment of
$n$ consecutive sites contains an even number of particles at time $t$. 
A site can be either empty or occupied by a single particle,
so the probability that a site is occupied, {\it i.e.}, the
particle density, is
\begin{equation}
\rho(t)=1-G_1(t)\;.
\end{equation}
Since the dynamic rules of the two-offspring BAW conserve parity, the only
way that
$G_n$ might change is when: ({\bf i})~particles at the edge of the segment hop
outside or branch, or ({\bf ii})~particles just outside of the segment hop
inside or branch. Let $\Gn$ and $\Fn$ represent segments
of even and odd number of particles, respectively. Then, the changes of
$G_n$ can be described schematically by the events:
\begin{eqnarray}
\label{schematic}
\frac{d}{dt}{G_n} & = & {\mathop{\Fn}^{n}{\shortstack {$\!\!\!\leftarrow$
\\
$\bullet$}}}  +
{\mathop{\Gn}^{n-1}{\shortstack {$\rightarrow\!\!$ \\
$\bullet$}}}  -
{\mathop{\Gn}^{n}{\shortstack {$\!\!\!\leftarrow$ \\
$\bullet$}}}  -
{\mathop{\Fn}^{n-1}{\shortstack {$\rightarrow\!\!$ \\
$\bullet$}}} \nonumber \\
& + & {\mathop{\Fn}^{n}\hspace*{-6pt}{\shortstack
{$\swarrow\hspace*{-6pt}|\hspace*{-6pt}\searrow$ \\
$\bullet$}}} +
 {\mathop{\Gn}^{n-1}\hspace*{-6pt}{\shortstack
{$\swarrow\hspace*{-6pt}|\hspace*{-6pt}\searrow$ \\
$\bullet$}}} -
 {\mathop{\Gn}^{n}\hspace*{-6pt}{\shortstack
{$\swarrow\hspace*{-6pt}|\hspace*{-6pt}\searrow$ \\
$\bullet$}}} -
 {\mathop{\Fn}^{n-1}\hspace*{-6pt}{\shortstack
{$\swarrow\hspace*{-6pt}|\hspace*{-6pt}\searrow$ \\
$\bullet$}}}\;.
\end{eqnarray}
Arrows in the first four terms indicate the hopping of a particle to the
left or right.  In the last four terms, the arrows indicate
branching. For example, the first term in~(\ref{schematic}) represents the
event that a particle outside of an
$n$-segment of odd parity jumps in, thus creating an $n$-segment of even
parity.

For the case of immediate reactions, $r=1$, the processes indicated
in~(\ref{schematic}) occur regardless of the state of target sites.  It is easy to show  
(see \cite{book,parity_ints,masser,lind}) that:
\begin{subequations}
\begin{eqnarray}
&&F_n\equiv\Pr(\mathop{\Gn}^n\bullet)=\case{1}{2}[(1-G_1)+(G_n-G_{n+1})]\;,\\
&&H_n\equiv\Pr(\mathop{\Fn}^n\bullet)=\case{1}{2}[(1-G_1)-(G_n-G_{n+1})]\;.
\end{eqnarray}
\end{subequations}
Using these relations, Eq.~(\ref{schematic}) becomes
\begin{equation}
\label{Gn.r1} {d\over dt}G_n(t)=
2(\Gamma+\Omega)(G_{n-1}-2G_n+G_{n+1})\;.
\end{equation}
$\Gamma$ and $\Omega$ are the rates of hopping and branching, respectively,
and the factor of 2 accounts for the events in~(\ref{schematic}) taking
place also at the left edge of the interval. The case of
$n=1$ requires a special equation, since
$G_0$ is undefined. Taking into account all the ways $G_1$ might change, we
find
\begin{equation}
\label{G1.r1} {\partial\over\partial t}G_1(t)=
2\Gamma(1-2G_1+G_2)+2\Omega(G_2-G_1)\;.
\end{equation}
Thus, Eq.~(\ref{Gn.r1}) may be understood to be valid also for $n=1$,
provided that one uses the boundary condition
\begin{equation}
\label{G0.bc} 
 G_0 = {{\Gamma + \Omega G_1}\over{\Gamma+\Omega}}
\;.  
\end{equation}
Additionally, since the $G_n$ are {\it probabilities\/}, we have
\begin{equation}
\label{Gn.bc} 0\leq G_n(t)\leq 1\;.
\end{equation}
Eq.~(\ref{Gn.r1}), with the boundary conditions~(\ref{G0.bc}),
(\ref{Gn.bc}) may be analyzed exactly for a variety of initial
conditions~\cite{parity_ints}.  Here we merely observe that the only steady state
solution supported by these equations is $G_n=1$, $n=1,2,\dots$,
corresponding to the absorbing state $\rhos=0$.
Indeed, the 2-BAW model with $r=1$ lacks an active phase and exhibits no transition.

\section{Parity intervals approximation for the general case \label{approxi}}
 
Consider now the case of finite reaction probability, $r<1$.  A
transition about some critical probability value $\rc$ is known to take
place, from the absorbing state into an active (non-empty) phase~\cite{2-BAW},
and our goal is to capture this transition, if only in an approximate
fashion.  We shall assume that $\Gamma=\Omega=1$, without loss of
generality~\cite{remark}.

Consider the first process on the r.h.s. of~(\ref{schematic}).  For $r<1$,
there is a difference in the reaction rate depending on whether
 the target site is empty or occupied:
\[
\mathop{\Fn}^{n}{\shortstack {$\!\!\!\leftarrow$\\ $\bullet$}}=
\mathop{\Fn}^{n-1}\circ{\shortstack {$\!\!\!\leftarrow$\\ $\bullet$}}+
r\,
\mathop{\Gn}^{n-1}\bullet{\shortstack {$\!\!\!\leftarrow$\\ $\bullet$}}\;.
\]
Unfortunately, events such as $\Fn\circ\bullet$ and
$\Gn\bullet\bullet$ cannot be expressed in closed form in terms of the
$G_n$.  We therefore rewrite the process in a way that the terms
associated with $r<1$, and which cannot be expressed in closed form, appear
as a perturbation, proportional to $s\equiv1-r$:
\begin{subequations}
\label{eq10}
\begin{eqnarray}
\mathop{\Fn}^{n}{\shortstack {$\!\!\!\leftarrow$\\ $\bullet$}} &&=
\mathop{\Fn}^{n-1}\circ{\shortstack {$\!\!\!\leftarrow$\\ $\bullet$}}+
r\,
\mathop{\Gn}^{n-1}\bullet{\shortstack {$\!\!\!\leftarrow$\\ $\bullet$}}
\nonumber\\
&&=
\mathop{\Fn}^{n-1}\circ\!\bullet+
\mathop{\Gn}^{n-1}\bullet\!\bullet-
(1-r)\,
\mathop{\Gn}^{n-1}\bullet\!\bullet\\
&&=
\mathop{\Fn}^{n}\bullet-s\,\mathop{\Gn}^{n-1}\bullet\!\bullet
\;.\nonumber
\end{eqnarray}
Likewise, the remainder of the diffusion events in~(\ref{schematic}) may
be rewritten as
\begin{eqnarray}
&&\mathop{\Gn}^{n-1}\shortstack{$\rightarrow\!\!$\\$\bullet$}=
  \mathop{\Gn}^{n-1}\bullet -s\,\mathop{\Gn}^{n-1}\bullet\bullet\;,\\
&&\mathop{\Gn}^{n}\shortstack{$\!\!\!\leftarrow$\\$\bullet$}=
  \mathop{\Gn}^{n}\bullet -s\,\mathop{\Fn}^{n-1}\bullet\bullet\;,\\
&&\mathop{\Fn}^{n-1}\shortstack{$\rightarrow\!\!$\\$\bullet$}=
  \mathop{\Fn}^{n-1}\bullet -s\,\mathop{\Fn}^{n-1}\bullet\!\bullet\;.
\end{eqnarray}
\end{subequations}
The branching events too require close inspection of the target sites. 
For example, we rewrite the first branching event of~(\ref{schematic}) in
a perturbative fashion:  
\begin{subequations}
\label{eq11}
\begin{eqnarray}
\mathop{\Fn}^{n}\hspace*{-6pt}{\shortstack
{$\swarrow\hspace*{-6pt}|\hspace*{-6pt}\searrow$ \\ $\bullet$}}&&=
\mathop{\Fn}^{n-1}\circ\hspace*{-5pt}{\shortstack
{$\swarrow\hspace*{-6pt}|\hspace*{-6pt}\searrow$ \\
$\bullet$}}\hspace*{-7pt}\circ+
r\,\mathop{\Gn}^{n-1}\bullet\hspace*{-5pt}{\shortstack
{$\swarrow\hspace*{-6pt}|\hspace*{-6pt}\searrow$ \\
$\bullet$}}\hspace*{-7pt}\circ+
r\,\mathop{\Fn}^{n-1}\circ\hspace*{-5pt}{\shortstack
{$\swarrow\hspace*{-6pt}|\hspace*{-6pt}\searrow$ \\
$\bullet$}}\hspace*{-7pt}\bullet+
r\,\mathop{\Gn}^{n-1}\bullet\hspace*{-5pt}{\shortstack
{$\swarrow\hspace*{-6pt}|\hspace*{-6pt}\searrow$ \\
$\bullet$}}\hspace*{-7pt}\bullet\nonumber\\
&&=(1-r)\,\mathop{\Fn}^{n-1}\circ\!\bullet\!\circ
+r\,\mathop{\Fn}^{n-1}\circ\!\bullet\!\circ
+r\,\mathop{\Gn}^{n-1}\bullet\!\bullet\!\circ
+r\,\mathop{\Fn}^{n-1}\circ\!\bullet\!\bullet
+r\,\mathop{\Gn}^{n-1}\bullet\!\bullet\!\bullet\\
&&=s\,\mathop{\Fn}^{n-1}\circ\!\bullet\!\circ
+(1-s)\,\mathop{\Fn}^n\bullet\;.\nonumber
\end{eqnarray}
The remainder of the branching terms are similarly expressed as
\begin{eqnarray}
&&\mathop{\Gn}^{n-1}\hspace*{-6pt}{\shortstack
{$\swarrow\hspace*{-6pt}|\hspace*{-6pt}\searrow$ \\ $\bullet$}}=
s\,\mathop{\Gn}^{n-2}\circ\!\bullet\!\circ
+(1-s)\,\mathop{\Gn}^{n-1}\bullet\;,\\
&&\mathop{\Gn}^{n}\hspace*{-6pt}{\shortstack
{$\swarrow\hspace*{-6pt}|\hspace*{-6pt}\searrow$ \\ $\bullet$}}=
s\,\mathop{\Gn}^{n-1}\circ\!\bullet\!\circ
+(1-s)\,\mathop{\Gn}^{n}\bullet\;,\\
&&\mathop{\Fn}^{n-1}\hspace*{-6pt}{\shortstack
{$\swarrow\hspace*{-6pt}|\hspace*{-6pt}\searrow$ \\ $\bullet$}}=
s\,\mathop{\Fn}^{n-2}\circ\!\bullet\!\circ
+(1-s)\,\mathop{\Fn}^{n-1}\bullet\;.
\end{eqnarray}
\end{subequations}
So far everything is exact.  In order to proceed, we approximate the terms
proportional to $s$ in the simplest possible way, by neglecting
correlations. Thus, for example, we write
\begin{equation}
\label{approx}
\Pr(\mathop{\Gn}^{n-1}\bullet\bullet)\approx
\Pr(\mathop{\Gn}^{n-1}\bullet)\Pr(\bullet)=F_{n-1}(1-G_1)\;,
\end{equation}
for the problematic term in~(\ref{eq10}a), and similar expressions for the
ones in (\ref{eq10}b)--(\ref{eq10}d).  For the problematic terms
proportional to $s$ in~(\ref{eq11}), we introduce the notation
\[
\Pr(\mathop{\Gn}^n\circ)=f_n\;,\qquad\Pr(\mathop{\Fn}^n\circ)=h_n\;,
\]
and
\[
\Pr(\bullet\bullet)=x\;,\qquad\Pr(\bullet\circ)=\Pr(\circ\bullet)=y\;,
\qquad\Pr(\circ\circ)=z\;,
\]
and approximate in the same spirit as above,
\begin{equation}
\Pr(\mathop{\Fn}^{n-1}\circ\bullet\circ)\approx
\Pr(\mathop{\Fn}^{n-1}\circ)\Pr(\bullet\circ)=h_{n-1}y\;,
\end{equation}
and similarly for the other terms.
Since $y+z=G_1$, $x+z=G_2$, and $x+2y+z=1$, it follows that
$x=\case{1}{2}(1-2G_1+G_2)$, $y=\case{1}{2}(1-G_2)$, and
$z=\case{1}{2}(-1+2G_1+G_2)$ are expressible in terms of the $G_n$ in
closed form.  Since $f_n+F_n=G_n$ and $h_n+H_n=1-G_1$, it follows that
$f_n=\case{1}{2}(-1+G_1+G_n+G_{n+1})$ and
$h_n=\case{1}{2}(1+G_1-G_n-G_{n+1})$ are too given in terms of the $G_n$ in
closed form.

We are now ready to write down a closed evolution equation for $G_n$. 
Starting from Eq.~(\ref{schematic}), using the representations of
Eqs.~(\ref{eq10}) and (\ref{eq11}) with their respective approximations,
collecting terms and rearranging, we obtain
\begin{equation}
\label{Gn.r}
\frac{d}{dt}G_n=\frac{1}{2}s(1-G_2)G_{n-2}+(2-3s+2sG_1)G_{n-1}
+\frac{1}{2}(-8+7s-4sG_1+sG_2)G_n+(2-s)G_{n+1}\;,
\end{equation}
valid for $n\geq3$.  As expected, this reduces to~(\ref{Gn.r1}), with
$\Gamma=\Omega=1$, in the limit $s\to0$.  The equation for
$n=2$ is exactly the same as~(\ref{Gn.r}), provided that one adopts the
boundary condition
\begin{equation}
\label{bcG0}
G_0=1\;.
\end{equation} 
$G_1$ requires a separate evolution equation that we find by
considering all the events that contribute to $dG_1/dt$, and write down,
in the spirit of Eq.~(\ref{Gn.r}), as
\begin{eqnarray}
\label{G1.r}
\frac{d}{dt}G_1&&=2(\bullet\shortstack{$\!\!\!\leftarrow$\\ $\bullet$}
+\shortstack{$\rightarrow\!\!$\\ $\bullet$} 
-\circ\shortstack{$\!\!\!\leftarrow$\\ $\bullet$}
+\bullet\hspace*{-6pt}\shortstack
{$\swarrow\hspace*{-6pt}|\hspace*{-6pt}\searrow$ \\ $\bullet$}
-\circ\hspace*{-6pt}\shortstack
{$\swarrow\hspace*{-6pt}|\hspace*{-6pt}\searrow$ \\ $\bullet$})
\nonumber\\
&&\approx 2\{(1-s)x+[(1-G_1)-sx]-y+(1-s)x-[(1-s)y+syG_1]\}\\
&&=2-6G_1+4G_2+s(-2+5G_1-4G_2+G_1G_2)\;.
\nonumber
\end{eqnarray}

To compute the steady state, we set $dG_n/dt=0$.  Eq.~(\ref{Gn.r}) yields
a recursion relation for the $G_n$, with constant coefficients (that
depend partly on $G_1$ and $G_2$).  Applying the ansatz $G_n=\l^n$ we
find a cubic equation for $\lambda$, with roots
\begin{equation}
\label{lambdas}
\l_0=1\;,\quad
\l_{\pm}=\frac{4-5s+4sG_1-sG_2\pm\sqrt{(4-5s+4sG_1-sG_2)^2+8s(2-s)(1-G_2)}}
{8-4s}\;.
\end{equation}
Thus $G_n$ has the general solution
\[
G_n=A\l_+^n+B\l_-^n+C\;,
\]
where $A$, $B$, $C$ are constants to be determined from boundary
conditions.  Since $0<s,G_1,G_2<1$ (in the active phase) it follows that
$|\l_{\pm}|<1$, and $C=\lim_{n\to\infty}G_n\equiv G_{\infty}$.  Suppose
that the initial distribution of particles is random, at density $\rho_0$,
then $G_n(t=0)=\case{1}{2}+\case{1}{2}(1-2\rho_0)^n$, and
$G_{\infty}(0)=1/2$~\cite{parity_ints}.  Because our model conserves parity, it
follows that $C=1/2$.  Furthermore, the boundary condition~(\ref{bcG0})
implies $B=\case{1}{2}-A$.  The remaining coefficient, $A$, could be
found from the relations
\begin{subequations}
\label{eq18}
\begin{eqnarray}
&&G_1=A\l_++(\case{1}{2}-A)\l_-+1/2\;,\\
&&G_2=A\l_+^2+(\case{1}{2}-A)\l_-^2+1/2\;,\\
&&G_2=\frac{2s-2+(6-5s)G_1}{4-4s+sG_1}\;.
\end{eqnarray}
\end{subequations}
The first two relations are required for self-consistence, since $\l_{\pm}$
depend on
$G_1$ and
$G_2$ (as well as on $s$).  Eq.~(\ref{eq18}c) is derived
from~(\ref{G1.r}), with $dG_1/dt=0$. Because Eqs.~(\ref{eq18}) are
unwielding, in practice we set a numerical value for $G_1$ and search for
$(A,s,G_2)$ that satisfy~(\ref{eq18}), using Mathematica.  We thus find a
kinetic phase transition which we next compare to simulations.

The critical point $\rc=1-\sc$, and the behavior of the order parameter
in its vicinity, can be obtained analytically, by perturbing~(\ref{eq18})
about the values $A=0$, $G_1=1$.  We thus find 
$ \rc=\frac{1}{6}\sqrt{33}-\frac{1}{2}=0.457427\dots $
and $ \rhos\sim(\rc-r)^{\beta}$, with $\beta=1\ $.  
The value of $\beta=1$ is in accordance with the expected from a mean-field like
solution.

We can also obtain analytically the behavior of the order parameter in the neighborhood
of $r=0$.  The answer, $\cs\sim1-\sqrt{2r}$, is the same as the one obtained from 
the one-site and two-site cluster approximations, and seems to fit
 the numerical data from simulations exactly.    Clearly, the role
 of fluctuations diminishes as $r$ decreases away from the critical
 point, and it is conceivable that their impact is negligible in the
 limit of zero reaction.

 \subsection*{Comparison to Simulations}

\begin{figure}
\centerline{\psfig{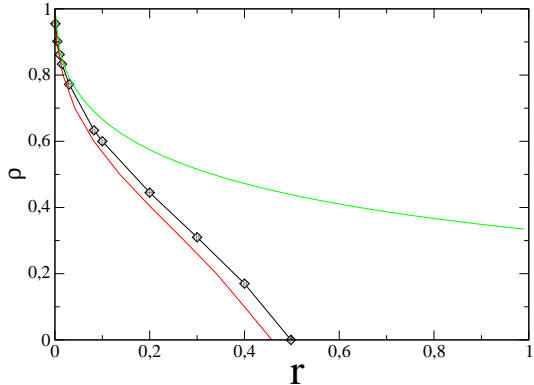}}
\vspace{0.5cm}
\caption{Order parameter $\rho$ as a function of $r$ as obtained from
(a)~Monte Carlo simulations (diamonds),  (b)~parity interval approximation 
(lowermost curve), and (c)~one-site cluster approximation (uppermost  curve):
the curve shows the stable root (see figure 1),
(observe that $\rho$ corresponds to  $c_{s}$ in the 
cluster approximation section).
Note the absence of a transition point in the third case.}
\label{Fig2}
\end{figure}

In order to test the limit of validity of the above calculations, 
we have performed a Monte Carlo simulation of the 2-BAW model. We use a system size 
$L=10^4$, and implement the previously discussed rules. 
The initial configuration consists of a randomly half-occupied lattice; 
we have verified that the obtained long-time asymptotic results are 
insensitive to variations of the initial condition. 
Plotting the average density as a function of time in a double logarithmic plot, 
we identify the critical point with the value of $r$ for which a separatrix,  
between curves that tend asymptotically to a constant (active phase) and curves that bend 
downwards (absorbing phase) converging asymptotically to a $t^{-1/2}$ decay \cite{exp} 
is obtained.
We thus find $r_c =0.495(10)$.  From the slope of the plot
the density decay critical exponent $\theta =0.28(1)$ (extending for about four
decades), is determined. 
It is in excellent agreement with previous measurements in the DP2 class \cite{Reviews}. 
The stationary density values are represented in Fig.~\ref{Fig2}. 

A point worth emphasizing is that finite-size corrections differ
from those usually encountered in systems with absorbing  states:
Rather than finding a larger stationary order parameter for smaller
systems, as is common, here one observes a faster decay to the absorbing phase.
Owing to this the exponent
ratio $\beta/\nu$ cannot be determined from finite-size scaling analysis. 

\begin{figure}
\centerline{\psfig{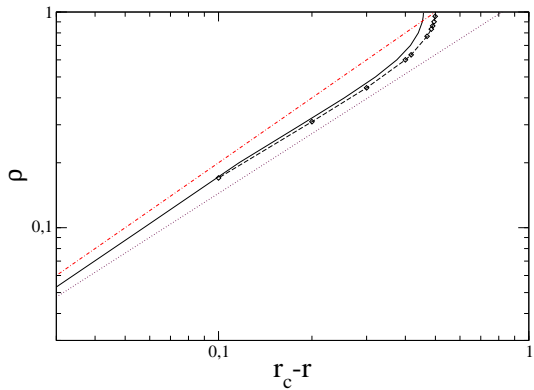}}
\vspace{0.5cm}
\caption{Log-log plot of the order parameter vs. the distance to the critical
 point. Shown are simulation results (curve with diamonds) compared to the
 prediction from the parity interval approximation (solid curve).  Slopes
 of 1 (top, dashed line) and 0.92 (bottom, dotted line) are shown for comparison. 
 The latter represents the accepted value of the order parameter exponent
 $\beta$ for the DP2 universality class in $d=1$.  Nearby criticality,  the
 mean-field solution converges to $\beta=1$.  Surprisingly, the mean-field
 slope agrees quite well with the expected value of 0.92, when $\rc-r\gtrsim0.1$.}
\label{Fig3}
\end{figure}

We have also measured the order-parameter critical exponent, by plotting the average 
stationary  density as a function of the distance to the critical point (Fig.~\ref{Fig3}).
Our finding of $\beta = 0.92(5)$ is again in good agreement with the commonly accepted value
\cite{Reviews}. Other exponents can certainly be measured, but with $\theta$ and $\beta$ we
can already guarantee that the model is indeed in the DP2 class.

\begin{figure}
\vspace{0.5cm}
\centerline{\psfig{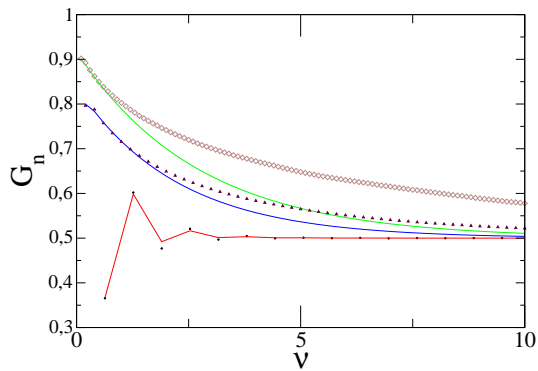}}
\vspace{0.5cm}
\caption{$G_n$ in the active phase for three different stationary
activity values: (from the bottom to the top) 
$0.633$, $0.2$, and $0.1$, as computed in a) Monte Carlo
simulation for $r=0.082$, $r=0.38$ and $r=0.45$ respectively (curves with symbols) 
and b) interval approximation (continuous lines). 
Instead of plotting as a function on $n$ in order to make more meaningful the comparison
 we use $\nu$, defined as $n$ divided the average
interparticle distance in each case.}
\label{Fig4}
\end{figure}

A more stringent test of the mean-field parity interval approximation is obtained
by comparing its predictions for $G_n$ to the numerical results of Monte Carlo
simulations, for different values of the control parameter $r$ (Fig.~\ref{Fig4}).
In all cases, we expect $G_{n}\to1/2$ for large $n$, since $G_{\infty}=1/2$ 
and parity is conserved.  The relevant question is how well the approximation
captures the $G_n$ for $n$ small.  Because the typical distance between
particles grows as one approaches criticality, we plot the results against
$\nu=\rho n$ rather than $n$, in order to compare them better.

For small values of $r$ (as shown in Fig.~\ref{Fig4}),
deep into  the active phase, spatial correlations play a minimal role and the 
theoretical prediction matches simulations remarkably well, for all $\nu$.
As $r$ is raised toward the critical point, correlations play a larger role and
the fit worsens. For example for $r=0.38$ ($\rho=0.2$) 
the theoretical curve fits experiments well only up to $\nu\approx 2$ (and, of
course, also at $\nu\gg1$).  Closer still to the critical point, at $r=0.45$
($\rho=0.1$), the theoretical prediction is good only up to $\nu\approx 1$.
  The large deviations observed for $\nu\gtrsim1$ in this last case indicate 
that the state of the system near the DP2 transition, despite being dilute,
 is highly ordered and correlated.

 The approximate curves, give converge to the exact value for $G_n \to 1/2$ 
for asymptotically large values of $\nu$.

\section{Discussion and future perspectives \label{disc}}
 
We have presented an approximation, based on the method of parity intervals, 
for the analysis of the  DP2 transition observed in the 2-BAW model in
one dimension with finite reaction rates. 
 The key ingredient of our approach is that it respects the parity conservation
implicit in the dynamical rules of the 2-BAW model: parity conservation
is here responsible for the $Z_2$ symmetry that underlies transitions
in the DP2 class, and provides a surprisingly good description of the 
transition at a mean-field level.

An interesting finding is the fact that spatial correlations play no role in
the limit of zero reaction rate, $r\to0$.  In this ``saturation" limit, the
dependence of the order parameter upon the critical field, $\rho\sim1-\sqrt{2r}$,
is captured even by the simplest one-site cluster approximation.

Extensive numerical simulations of the DP2 transition suggest a value of
$\beta=0.92(2)$  for the order parameter critical exponent \cite{Reviews}.
On the other hand, an exact value $\beta=1$ was conjectured by Jensen 
and later supported by numerical simulations combined with 
Pad\'e approximants \cite{Inui}. 
Regrettably, we have little to add to this controversy.  
Our mean-field results away from criticality ($\rc-r\gtrsim0.1$) are 
consistent with $\beta=0.92$, while $\beta\to1$ as we approach the 
transition (Fig.~\ref{Fig3}).  However, the large spatial
correlations observed near criticality are not faithfully modeled by our
mean-field like approach (Fig.~\ref{Fig4}), and the values we obtain for
$\rc-r\lesssim0.1$ are unreliable.  
It would be desirable to improve our approximation so as to better model 
regions closer to $\rc$.  Our attempts
in this direction have been futile, so far. 
Arguably, the simplest improvement
would be to replace the approximation of Eq.~(\ref{approx}) by the more accurate:
\[
\Pr(\mathop{\Gn}^{n-1}\bullet\bullet)\approx
\frac{\Pr(\mathop{\Gn}\limits^{n-1}\bullet)\Pr(\bullet\bullet)}{\Pr(\bullet)}\;,
\]
and likewise elsewhere.  However, on employing this scheme, instead of an improved
 result we find that the transition disappears.   A similar phenomenon is known to
 occur also in mean-field cluster approximations, where sometimes increasing the
 cluster size yields lower quality predictions.

Other interesting open prospects include using the parity interval approximation
 for the analysis of dynamical aspects of the DP2 transition.  We have here
 systematically assumed that all time derivatives are zero, thereby accessing
 the steady state alone.  Analyzing the very same equations with the full
 time dependence built in should yield a prediction for the order parameter
 time decay.  Additionally, the equations could be modified to describe a
 spatially inhomogeneous system.  In that case, one could study the dynamics of spreading.

\vspace{0.5cm}

\acknowledgments 
We acknowledge  financial support from the Spanish MCyT (FEDER) under project BFM2001-2841.
DbA thanks the NSF for partial support, under contract no.\ PHY-0140094,
 as well as the warm hospitality of the Instituto de F{\'\i}sica Te{\'o}rica 
y Computacional Carlos I, University of Granada, during a crucial phase of the project.
We also gratefully thank G. Odor and A. Szolnoki, for very useful comments on the cluster
mean field calculations that have helped to improve the paper.


\begin{thebibliography}{29}
\bibitem{remark}  The same transition is observed whether one fixes $r$
and treats $\Gamma/\Omega$ as a critical field, or whether one fixes
$\Gamma/\Omega$ and takes $r$ to be the critical field.  Here we adopt the
latter, with $\Gamma/\Omega=1$.
\bibitem{exp} Observe that in
this type of systems we do not have an exponential decay in the absorbing phase, 
but a power-law one, controlled asymptotically by the reaction $A + A \to 0$.
\end{thebibliography}
\end{document}